\documentclass[twocolumn,pra,twocolumn,superscriptaddress]{revtex4}
\usepackage[latin9]{inputenc}
\usepackage{color}
\usepackage{amsmath}
\usepackage{amssymb}
\usepackage{graphicx}
\usepackage{tikz}
\usepackage{mathrsfs} 
\usepackage{float}
\usepackage{bbm}
\usepackage{epstopdf}
\usepackage{epsfig}
\usepackage{verbatim}
\usepackage{array}
\usepackage{setspace}
\usepackage{times}
\usepackage{tabularx}
\usepackage{setspace}
\usepackage{ulem}

\usepackage[unicode=true,
bookmarks=true,bookmarksnumbered=false,bookmarksopen=false,
breaklinks=false,pdfborder={0 0 1},backref=false,colorlinks=true]
{hyperref}
\hypersetup{
	linkcolor=magenta, urlcolor=blue, citecolor=blue, pdfstartview={FitH}, hyperfootnotes=false, unicode=true}

\newcolumntype{C}[1]{>{\centering\arraybackslash$}p{#1}<{$}}

\begin{document}

\title{Universal singlet-triplet qubits implemented near the transverse sweet spot}

\author{Wen-Xin Xie}
\affiliation{Guangdong Provincial Key Laboratory of Quantum Engineering and Quantum Materials,
	and School of Physics\\ and Telecommunication Engineering, South China Normal University, Guangzhou  510006, China}

\author{Chengxian Zhang}\email{cxzhang9-c@my.cityu.edu.hk}
\affiliation{Guangdong Provincial Key Laboratory of Quantum Engineering and Quantum Materials,
	and School of Physics\\ and Telecommunication Engineering, South China Normal University, Guangzhou  510006, China}
\affiliation{Guangdong-Hong Kong Joint Laboratory of Quantum Matter and  Frontier Research Institute for Physics,\\ South China Normal University, Guangzhou  510006, China}

\author{Zheng-Yuan Xue}  \email{zyxue@scnu.edu.cn}
\affiliation{Guangdong Provincial Key Laboratory of Quantum Engineering and Quantum Materials,
	and School of Physics\\ and Telecommunication Engineering, South China Normal University, Guangzhou  510006, China}
\affiliation{Guangdong-Hong Kong Joint Laboratory of Quantum Matter and  Frontier Research Institute for Physics,\\ South China Normal University, Guangzhou  510006, China}

\date{\today}

\begin{abstract}

The key to realizing fault-tolerant quantum computation for singlet-triplet (ST) qubits in semiconductor double quantum dot (DQD) is to operate both the single- and two-qubit gates with high fidelity. The feasible way includes operating the qubit near the transverse sweet spot (TSS) to reduce the leading order of the noise, as well as adopting the proper pulse sequences which are immune to noise. The single-qubit gates can be achieved by introducing an AC drive on the detuning near the TSS. The large dipole moment of the DQDs at the TSS has enabled strong coupling between the qubits and the cavity resonator, which leads to a two-qubit entangling gates. When operating in the proper region and applying modest pulse sequences, both single- and two-qubit gates are having fidelity higher than 99\%. Our results suggest that taking advantage of the appropriate pulse sequences near the TSS can be effective to obtain high-fidelity ST qubits.

\end{abstract}

\maketitle

\section{Introduction}

Quantum computing using the spin states of the electrons confined in the semiconductor quantum dots \cite{Loss.98,Petta.05,Hanson.07,Zimmerman.14,Bermeister.14,Veldhorst.14, Veldhorst.15,Takedae.16,Schreiber.18, Watson.18,Yoneda.18,Chan.18,sigillito.19,Crippa.19,Emerson.19,Huang.19} is promising due to the long coherence time and the possibility for scalability \cite{Hanson.07}. Singlet-triplet (ST) qubits in double quantum dot (DQD) \cite{Petta.05,foletti.09,wang.14,shulman.14,nichol.17,zhang.17,takeda.20,cerfontaine.20} is particularly standing out of various proposals that are encoded using the spin states because of its all-electrical operation and strong exchange interaction. However, in the operation for the ST qubits, both single- and two-qubit gates are susceptible to the charge fluctuation (charge noise). Although recent experiments have shown that the single-qubit gate fidelity for the ST qubits can be higher than 99\% \cite{nichol.17,cerfontaine.20,takeda.20}, the fidelity for the two-qubit gate reported until now has been lower than the requirement for the fault-tolerant quantum computing \cite{shulman.12,nichol.17}. This motivates us to further search for useful methods to suppress the charge noise. 

Recently, much works have focused on the symmetric operation of the sweet spot (SOP) \cite{Medford.13,Shim.16,Russ.16,Marko.17,Zhang.18,Yang.19}, which is far away from the charging transition regions, namely (1, 1) to (0, 2) or (2, 0), where the notation $(n_{L},n_{R})$ denotes the number of electron in the left and right dot. By operating the qubit near the sweet spot, the derivative of the exchange energy with respect to detuning is minimized. Therefore, the sensitivity of the qubits to the leading order of charge noise is reduced. In the SOP region, the dipole moment is dominated by the longitudinal components, which can be used to construct the CPHASE two-qubit gate via the longitudinal coupling \cite{harvey.18}. On the other hand, the transverse dipole moment is very small and thus the transverse coupling is infeasible. A latest work \cite{abadillo.19} has suggested that there exists a family of the so-called transverse sweet spots (TSS). It is shown that even though TSS are far from the conventional SOP, they are still having the merit of the sweet spot as the same with the SOP. Near the TSS region, the transverse dipole moment is large while the longitudinal one is suppressed, which is contrary to the case for the SOP. Therefore, they have enabled strong transverse couplings. This is promising to design a two-qubit iSWAP gate by transversely coupling two qubits mediated by a cavity resonator, as shown in the recent works \cite{blais.07,srinivasa.16,harvey.18,abadillo.19,zhang.20a,zhang.20b}.

In this work, we have investigated the ST qubits under the realistic time-dependent noise environment in DQD. By introducing an AC-driven detuning near the TSS, one is able to achieve arbitrary single-qubit gates. In addition, we have studied the performance of the gates in two different regions, which corresponds to the TSS value of $\varepsilon_{\rm{SS}}>0$ and $\varepsilon_{\rm{SS}}<0$, respectively. Here, $\varepsilon_{\rm{SS}}$ is the detuning at TSS. When operating the gate in the region $\varepsilon_{\rm{SS}}>0$, the leakage is minimized and the detuning noise remains only in the $z$ direction. We compare the naive (primitive) gate with several typical pulse sequences, namely, the non-cyclic geometric gate \cite{liu.20}, the cyclic geometric gate \cite{Zhao.17,zhang.20a} and the CORPSE gate \cite{Cummins.03,Bando.13}, which are well-known to suppress the $z$-component noise. We find that the gate performance under the time-dependent noise for each type of gate has strong connection with the noise spectrum. Whether the naive gates can be improved sensitively depends on the chosen rotations.  On the other hand, we have also found that using the strong dipole coupling between the DQDs and the cavity resonator, a two-qubit entangling gate can be achieved. In our work, the fidelity for both single- and two-qubit gates are surpassing 99\%.

\section{Model}\label{sec:model}

\begin{figure}
	\includegraphics[width=0.92\columnwidth]{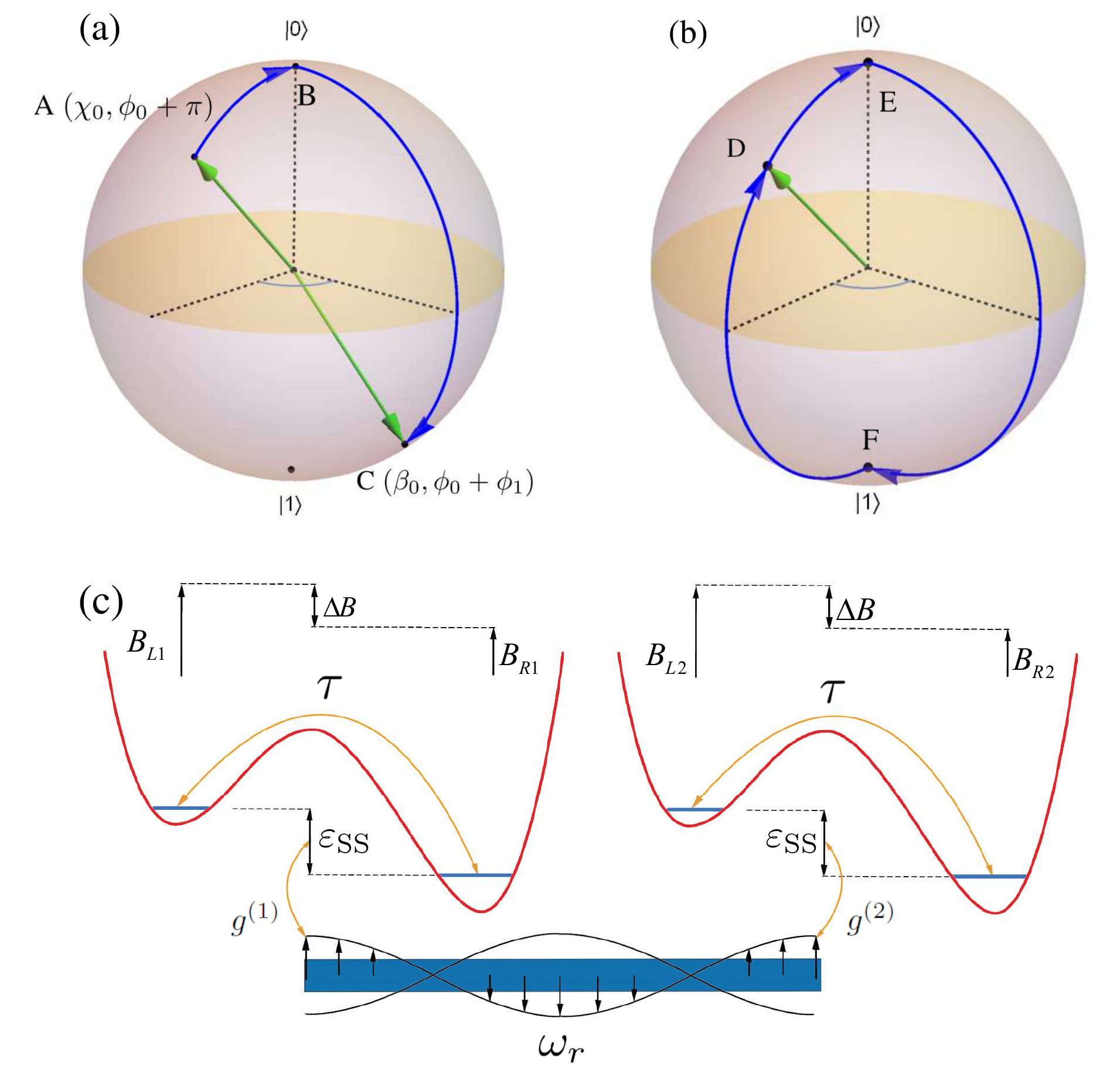}
	\caption{(a) The evolution path of state $\left| {{\psi _ + }} \right\rangle$, along the non-cyclic path A-B-C to construct the non-cyclic geometric gate. (b) The evolution path of state $\left| {{\psi _ + }} \right\rangle$, along the cyclic path D-E-F-D so as to obtain the geometric gate. (c) Dipole coupling of two DQDs to the cavity resonator. For both single- and two-qubit  gates, the detuning is operated at the TSS denoted as $\varepsilon_{\rm{SS}}$.}
	\label{fig:model}
\end{figure}

\begin{figure}
	\includegraphics[width=1.\columnwidth]{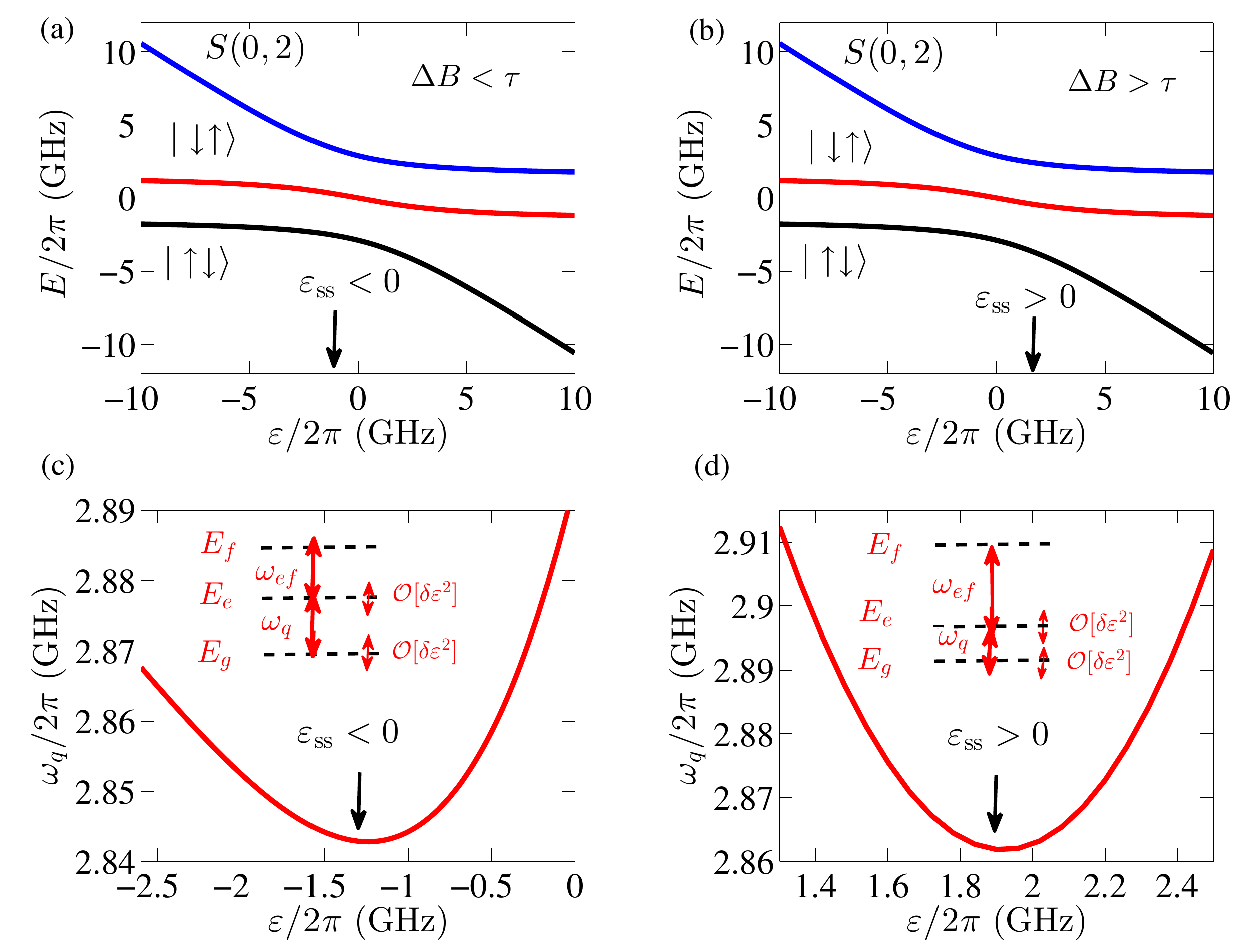}
	\caption{(a) and (b): Energy spectrum of the DQD Hamiltonian $H_{\rm{ST}}$ in Eq.~\ref{eq:hst} as a function of the detuning value $\varepsilon$ considering the cases $\Delta B<\tau$ (a) and $\Delta B>\tau$ (b). (c) and (d): The corresponding qubit energy $\omega_{q}=E_{e}-E_{g}$ is drawn from (a) and (b), respectively. When $\Delta B<\tau$, the TSS point appears at $\varepsilon<0$. For comparison, it appears at $\varepsilon>0$ when $\Delta B>\tau$. Parameters for (a) and (c): $\Delta B/2\pi=1.5\ \rm{GHz}$, $\tau/2\pi=1.75\ \rm{GHz}$ while for  (b) and (d): $\Delta B/2\pi=2.5\ \rm{GHz}$, $\tau/2\pi=1.5\ \rm{GHz}$. }
	\label{fig:energy}
\end{figure}
The Hamiltonian for the ST qubits in the basis states spanned by $\left\{\left|T_{0}(1,1)\right\rangle,|S(1,1)\rangle,|S(0,2)\rangle\right\}$ is given by \cite{abadillo.19}
\begin{equation}
H_{\rm{ST}}=\left(\begin{array}{ccc}
0 & \Delta B & 0 \\
\Delta B & 0 & \sqrt{2} \tau \\
0 & \sqrt{2} \tau & -\varepsilon
\end{array}\right).
\label{eq:hst}
\end{equation}
Here, the spin state is defined as $\left|\uparrow,\downarrow\right\rangle=c_{L \uparrow}^{\dagger} c_{R \downarrow}^{\dagger}|\mathcal{V}\rangle$ where $c_{j \mu}^{\dagger}$ creates an electron with spin $\mu$ at the $j$th quantum dot and $\mathcal{V}$ denotes vacuum \cite{wang.14}. $|S(1,1)\rangle=(\left|\uparrow,\downarrow\right\rangle-\left|\downarrow,\uparrow\right\rangle) / \sqrt{2}$ and
$|T_{0}(1,1)\rangle=(\left|\uparrow, \downarrow\right\rangle+\left|\downarrow, \uparrow\right\rangle) / \sqrt{2}$
are the spin singlet and triplet state, respectively. While $|S(0,2)\rangle$ is another singlet state when both of the two electrons occupy the right dot.   The parameter $\Delta B$ represents the magnetic field gradient between the two quantum dots, while $\tau$ and $\varepsilon$ are the tunneling and the detuning values between the two dots (see Fig.~\ref{fig:model}(c)). Note that the parameters here are in energy units by taking $\hbar=1$. Also, we assume the global Zeeman field is large enough, therefore, the other two triplet states $\left|\uparrow ,\uparrow\right\rangle$ and $\left|\downarrow, \downarrow\right\rangle$ are split off and thus can be ignored.

We first show how to seek for the proper TSS detuning point $\varepsilon_{\rm{SS}}$ to define the qubit subspace. As shown in Fig.~\ref{fig:energy}, the energy structure of the system is plotted by diagonalizing $H_{\rm{ST}}$, where the eigenbasis is denoted as $\{|g\rangle,|e\rangle,|f\rangle\}$. The qubit states are defined as the two lowest eigenbasis states, namely $\left\{|0\rangle=|e\rangle,|1\rangle=|g\rangle\right\}$. Introducing an AC drive $\varepsilon'(t)=\varepsilon_{\rm{AC}}\cos(\omega t+\phi(t))$ to the detuning parameter oscillating near the given $\varepsilon_{\rm{SS}}$, namely, $\varepsilon=\varepsilon_{\rm{SS}}+\varepsilon'$, the total Hamiltonian of the qubit system is therefore
\begin{equation}
H_{q}=\sum_{n=1}^{3} E_{n}(\varepsilon_{\rm{SS}}) \sigma_{n n}+H_{\rm{int}}
\label{eq:hqubit},
\end{equation}
with
\begin{equation}
\begin{aligned}
H_{\rm{int}}=-\varepsilon'|S(0,2)\rangle\langle S(0,2)|=\varepsilon' \sum_{m, n} d_{mn} \sigma_{m n}.
\end{aligned}
\label{eq:hint}.
\end{equation}
Here, $E_{n}(\varepsilon_{\rm{SS}})$ (the numbers 1, 2, and 3 represent states $|g\rangle$, $|e\rangle$ and $|f\rangle$) is the eigenvalue of $H_{\rm{ST}}$ which is dependent on the chosen $\varepsilon_{\rm{SS}}$, $\sigma_{mn}=|m\rangle\langle n|$ with $m, n$ being the eigenstates, and $d_{mn}=\langle m|\hat{d}| n\rangle$ is the dipole matrix element with $\hat{d}=\partial H_{\mathrm{ST}} / \partial \varepsilon$. When the AC drive is in resonance with the qubit energy, i.e. $\omega_{q}=E_{e}-E_{g}=\omega$, the system under the rotating wave approximation (RWA) in the rotating frame can be well described by an effective two-level structure (the detail can be seen in Appendix.~\ref{appx:Heff}):
\begin{equation}
H_{\rm{AC}}=\frac{\Omega_{0}}{2}\left(\cos \phi(t)\ \sigma_{x}+\sin \phi(t)\ \sigma_{y}\right)
\label{eq:hac},
\end{equation}
Here, the Rabi frequency is $\Omega_{0}=|d_{ge}|\varepsilon_{\rm{AC}}$. Then, using this piecewise continuous Hamiltonian one is able to realize arbitrary rotation around the axis on the $x$-$y$ plane, denoted as $R(\boldsymbol{r},\theta)$ where $\boldsymbol{r}$ is the rotation axis and $\theta$ the rotation angle. The rotation suffering noise is as described in Appendix.~\ref{appx:fidelity}. Further, Any single-qubit gate out of the $x$-$y$ plane can be decomposed into a related ``$x$-$y$-$x$'' (or ``$y$-$x$-$y$'') sequence \cite{nielsen.02}, which we call as ``naive pulse sequence''. 

\begin{figure}
	\includegraphics[width=1\columnwidth]{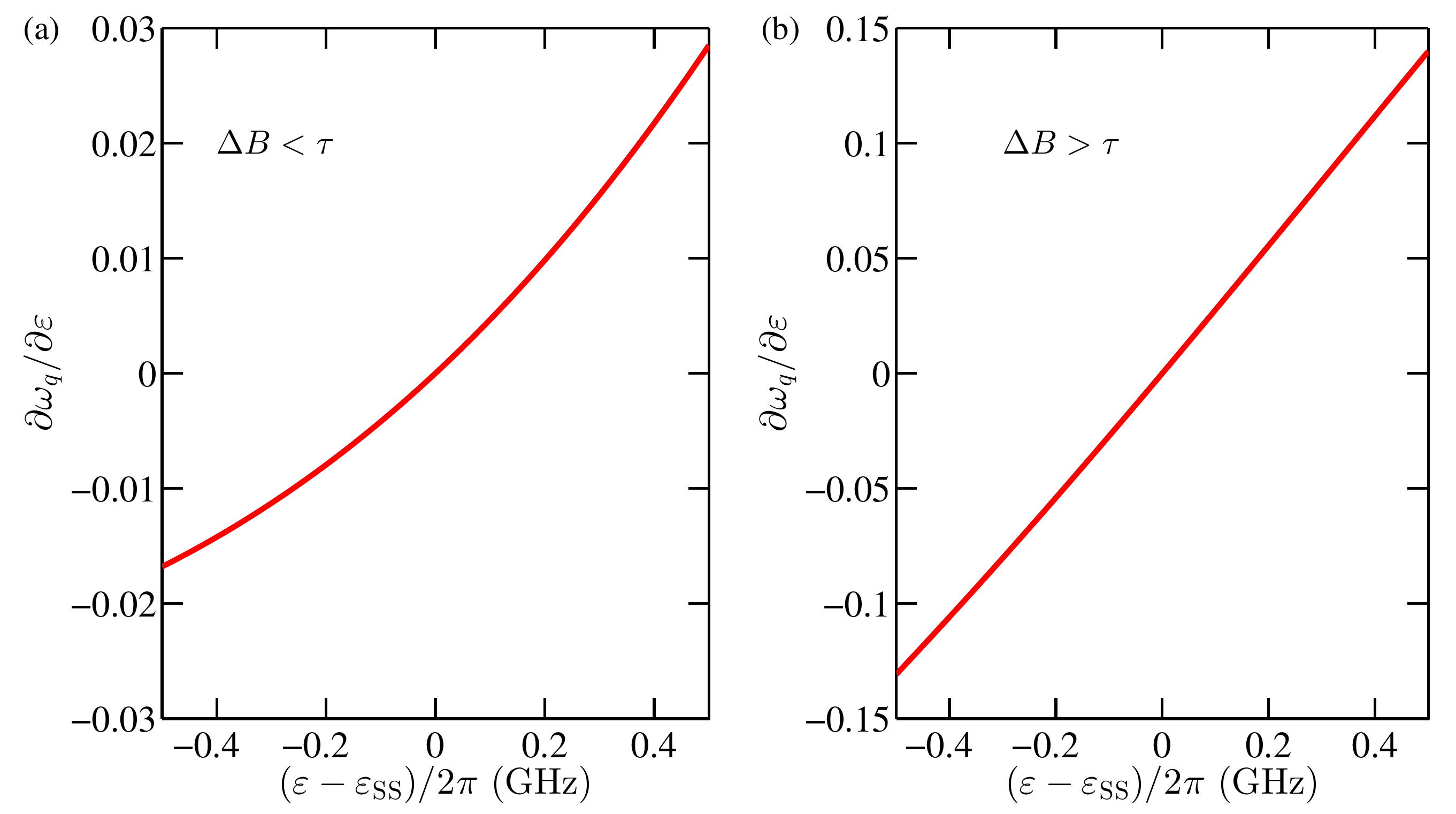}
	\caption{The derivative $\partial{\omega_{q}}/\partial{\varepsilon}$ around the TSS. Parameters for (a): $\Delta B/2\pi=1.5\ \rm{GHz}$, $\tau/2\pi=1.75\ \rm{GHz}$ while for (b): $\Delta B/2\pi=2.5\ \rm{GHz}$, $\tau/2\pi=1.5\ \rm{GHz}$.}
	\label{fig:derivative}
\end{figure}

Normally, $H_{\rm{int}}$ can induce the transverse and longitudinal coupling of the qubit. At the TSS, $|d_{ge}|$ is closed to the maximum value (about 0.5) and the longitudinal component is small which can be further suppressed because it appears as the fast-oscillating terms. Therefore, it is wise to operate the qubit near the TSS. On the other hand, the TSS can be found in two different regions, i.e. $\Delta B<\tau$ and $\Delta B>\tau$. When $\Delta B<\tau$, the corresponding detuning value of the TSS is negative, i.e. $\varepsilon_{\rm{SS}}<0$ (see Fig.~\ref{fig:energy}(a) and (c)). In this case, we have $\omega_{ef}=E_{f}-E_{e}\simeq\omega_{q}$ and it is easy to introduce leakage to the excited state $|f\rangle$. The mechanism of the leakage effect is described in Appendix.~\ref{appx:Heff}. For comparison, when $\Delta B>\tau$, we have $\varepsilon_{\rm{SS}}>0$ (see Fig.~\ref{fig:energy}(b) and (d)), where $\omega_{ef}$ is substantially larger than $\omega_{q}$ and therefore the leakage is suppressed. To strongly suppress the noise effect, we consider implementing geometric gate in the region $\Delta B>\tau$ and using $\varepsilon_{\rm{AC}}/2\pi=0.1\ \rm{GHz}$ in this work, which is reasonable in the experiment \cite{Kim.15}. Hereafter, we stick to the parameters: $\Delta B/2\pi=2.5\ \rm{GHz}$, $\tau/2\pi=1.5\ \rm{GHz}$. The corresponding TSS is $\varepsilon_{\rm{SS}}/2\pi=1.91935\ \rm{GHz}$ and the dipole element is $|d_{ge}|=0.45055$. Considering $\varepsilon_{\rm{AC}}/2\pi=0.1\ \rm{GHz}$, the Rabi frequency is thus $\Omega_{0}/2\pi=0.045055\ \rm{GHz}$.

\section{Results}\label{sec:results}
\subsection{Single-qubit gates}\label{sec:singlegeo}

Here, we focus on introducing how to design the non-cyclic geometric gate using the piecewise continuous Hamiltonian \cite{liu.20} (the other types of pulse sequences is introduced in Appendix.~\ref{appx:fidelity}), and then we will discuss the noise effect caused by the charge noise in the DQD.

The quantum gate considered here is using two-pieces of Hamiltonian in Eq.~\ref{eq:hac}. In each of the piece, the Hamiltonian is with the form as 
\begin{widetext}
\begin{equation}
H_{n}(t)=\left\{\begin{array}{ll}
\frac{\Omega_{0}}{2}(\cos (\phi_{0} + \frac{\pi }{2})\ \sigma_{x}+\sin (\phi _{0} + \frac{\pi }{2})\ \sigma_{y}), & T_{A} \leqslant t< T_{B} \\
\frac{\Omega_{0}}{2}(\cos (\phi_{0} + \phi_{1} + \frac{\pi }{2})\ \sigma_{x}+\sin (\phi _{0} + \phi _{1} + \frac{\pi }{2})\ \sigma_{y}), & T_{B} < t\leqslant T_{C}
\end{array}\right.
\label{eq:paths}
\end{equation}
\end{widetext}
where $T_{B}-T_{A}=\chi_{0}/\Omega_{0}$ and $T_{C}-T_{B}=\beta_{0}/\Omega_{0}$. Then, this two-connecting evolution together at the final time forms a desired quantum gate as
\begin{widetext}
\begin{equation}
\begin{aligned}
{U_{n}}\left( {{\chi _0},{\phi _{0}},{\phi _{\rm{1}}},{\beta _0}} \right)
&={U_{CB}}\left( {{T_{C}},{T_{B}}} \right){U_{BA}}\left( {{T_{B}},T_{A}} \right)\\
&=\left(\begin{array}{ccc}
\cos \frac{\chi_{0}}{2} \cos \frac{\beta_{0}}{2}-\sin \frac{\chi_{0}}{2} \sin \frac{\beta_{0}}{2} e^{-i \phi_{1}} & -\cos \frac{\chi_{0}}{2} \sin \frac{\beta_{0}}{2} e^{-i\left(\phi_{0}+\phi_{1}\right)}-\cos \frac{\beta_{0}}{2} \sin \frac{\chi_{0}}{2} e^{-i \phi_{0}} \\
\cos \frac{\chi_{0}}{2} \sin \frac{\beta_{0}}{2} e^{i\left(\phi_{0}+\phi_{1}\right)}+\cos \frac{\beta_{0}}{2} \sin \frac{\chi_{0}}{2} e^{i \phi_{0}} & \cos \frac{\chi_{0}}{2} \cos \frac{\beta_{0}}{2}-\sin \frac{\chi_{0}}{2} \sin \frac{\beta_{0}}{2} e^{i \phi_{1}}
\end{array}\right)
\end{aligned}
\label{eq:noncy}
\end{equation}
\end{widetext}
Here, $\chi_{0}$, $\phi_{0}$, $\phi_{1}$ and $\beta_{0}$ are determined by the chosen evolution operator. In our simulation, the specific parameters are present in Appendix.~\ref{appx:table}. Actually, $ {U_{n}}\left( {{\chi _0},{\phi _{\rm{0}}},{\phi _{\rm{1}}},{\beta _0}} \right) $ can enable arbitrary rotation on the Bloch sphere. For example,  $ {U_{n}}({\chi _0},\frac{\pi }{2},\pi ,{\beta _0})$  and ${U_{n}}({\chi _0}, - \pi ,\pi ,{\beta _0})$ can implement a rotation around $x$ axis and $y$ axis, respectively, the rotation angle for both of which have the same form as $\beta_{0}-\chi_{0}$. Note that, for a desired rotation around the  $x$ or $y$ axis, even though one has many choices to pick up $\chi_{0}$ and $\beta_{0}$, the conditions for them are clear: $\chi_{0}>0$ and $\beta_{0}>0$. In this way, the evolution time is positive and it ensures the gate will not degenerate to the naive pulse sequence.

The evolution of the quantum gate can be visualized by using the orthogonal dressed states
\begin{equation}
\begin{aligned}
\left|\psi_{+}(t)\right\rangle=\cos \frac{\chi}{2} e^{-\frac{i}{2} \eta}|0\rangle+\sin \frac{\chi}{2} e^{\frac{i}{2}  \eta}|1\rangle \\
\left|\psi_{-}(t)\right\rangle=\sin \frac{\chi}{2} e^{-\frac{i}{2}  \eta}|0\rangle-\cos \frac{\chi}{2} e^{\frac{i}{2}  \eta}|1\rangle
\end{aligned}
\end{equation}
As shown in Fig.~\ref{fig:model}(a), the whole evolution path of the quantum gate is divided into two proper parts, which are denoted as path AB and BC. In the first part of the evolution, $\left|\psi_{+}\right\rangle $ starts from a given point A, and evolves along the geodesic line up to the north pole B. Then, in the second part, it goes down to a given point C along another geodesic line. Using this dressed states basis, the evolution operator $ {U_{n}}\left( {{\chi _0},{\phi _{\rm{0}}},{\phi _{\rm{1}}},{\beta _0}} \right) $ can also be written as
\begin{equation}
\begin{aligned}
{U_{n}} = {e^{i{\gamma}}}\left| {{\psi _+}({T_{C}})} \right\rangle \left\langle {{\psi _+}({T_{A}})} \right| + {e^{ - i{\gamma}}}\left| {{\psi _{-}}({T_{C}})} \right\rangle \left\langle {{\psi _{-}}({T_{A}})} \right|
\end{aligned}
\end{equation}
where $\gamma$ is the related phase that is obtained. This two-piece path in the dressed state basis is not closed. On the other hand, the evolution path in the dressed state basis can alternatively be cyclic, as shown in Fig.~\ref{fig:model}(b) which can form a geometric gate. The detail of the geometric gate is as shown in Ref.~\cite{Zhao.17}. For this geometric gate, its whole evolution time is fixed to be $2 \pi/\Omega_{0}$ \cite{zhang.pra}. By carefully selecting proper $\chi_{0}$ and $\beta_{0}$ so as to satisfy $\chi_{0}+\beta_{0}<2\pi$, the evolution time for the non-cyclic geometric gate can be always shorter than the geometric ones. Note that, in Ref.\cite{liu.20}, the authors have pointed out that the non-cyclic geometric gate introduced here is non-cyclic and non-Abelian. However, for the case $m\neq n$ with $m,n \in +,-$, one finds $\int_{0}^{\tau}\left\langle\psi_{m}(t)|\mathcal{H}(t)| \psi_{n}(t)\right\rangle d t\neq0$, namely, the dynamical phase in this case is present. Therefore, it does not satisfy the parallel transport condition and cannot be the non-Abelian geometric gate \cite{kult.06}.

Then, we study the noise effect of this AC-driven system to the non-cyclic geometric  gate. Considering the charge noise effect, the detuning value turns to be $\varepsilon(t)=\varepsilon_{\rm{SS}}+\varepsilon_{\rm{AC}}\cos(\omega t+\phi(t))+\delta\varepsilon(t)$, where $\delta\varepsilon(t)$ is the time-dependent charge noise. $\delta\varepsilon(t)$ results in the error for the chosen $\varepsilon_{\rm{SS}}$, i.e. $\varepsilon_{\rm{SS}}\rightarrow\varepsilon_{\rm{SS}}+\delta\varepsilon(t)$ and further induce the drift for the energy-level structure, namely, $E_{n}\rightarrow E_{n}+\delta E_{n}$. Assuming that $\varepsilon_{\rm{AC}}$, $\delta\varepsilon(t)\ll \varepsilon_{\rm{SS}}$, we can expand the qubit energy as
\begin{equation}
\begin{aligned}
\omega_{q}\approx \omega_{q}\left(\varepsilon_{\rm{SS}}\right)+\delta\omega_{q},
\end{aligned}
\label{eq:expand}
\end{equation}
where we have $\delta\omega_{q}\simeq (\partial{\omega_{q}}/\partial{\varepsilon})\ \delta\varepsilon$ when operating near the TSS point. This in turn leads to error for $H_{\rm{AC}}$ in the rotating frame (see Appendix.~\ref{appx:Heff}):
\begin{equation}
H_{\rm{AC}}'=\frac{\Omega_{0}}{2}\left(\cos \phi(t)\ \sigma_{x}+\sin \phi(t)\ \sigma_{y}\right)+ \delta\omega_{q}/2\ \sigma_{z}
\label{eq:hac2}.
\end{equation}
Note that here we have assumed $\delta\varepsilon$ is a constant value for simplicity and its time-dependent effect will be considered later. At the TSS point the qubit is first-order insensitive to the charge noise because of $\partial{\omega_{q}}/\partial{\varepsilon}=0$. However, when it is away from the TSS point, this first-order charge noise effect cannot be ignored anymore. In Fig.~\ref{fig:derivative}, we show the derivative $\partial{\omega_{q}}/\partial{\varepsilon}$ as a function of the detuning value $\varepsilon$ around the TSS point. We find that the derivative in Fig.~\ref{fig:derivative}(a), which corresponds to $\Delta B<\tau$, is
much smaller than that one when $\Delta B>\tau$, as shown in Fig.~\ref{fig:derivative}(b). This means the qubit is more insensitive to the charge noise when operating in this region. However, as stated above, the leakage effect in this region is substantially large regardless of this superiority to the charge noise. Meanwhile, when $\Delta B>\tau$, $\partial{\omega_{q}}/\partial{\varepsilon}$ grows almost linearly with $\varepsilon-\varepsilon_{\rm{SS}}$. This implies that small AC drive is more appropriate in order to avoid large charge noise. As stated above, we have taken $\varepsilon_{\rm{AC}}/2\pi=0.1\ \rm{GHz}$.
Considering all the operating region of the detuning value $\varepsilon$ we have $\partial{\omega_{q}}/\partial{\varepsilon}\le \partial{\omega_{q}}/\partial{\varepsilon}|_{\varepsilon=\varepsilon_{\rm{AC}}}\simeq0.02$. According to Ref.~\cite{mielke.20}, the standard deviation related to the detuning charge noise is with the order of $\sigma_{\varepsilon}=1\ \mu \rm{eV}$. Thus, the deviation with respect to the qubit energy can be $0<\sigma_{\omega q}\leq(\partial{\omega_{q}}/\partial{\varepsilon})\ \sigma_{\varepsilon}\simeq0.02\ \mu \rm{eV}$. In our simulation we have used $\sigma_{\omega q}=0.02\ \mu \rm{eV}$.

Note that, for the ST qubits in semiconductor quantum dot, there are several extra noise channels that should be carefully treated. For example, the charge noise can also bring fluctuation for the tunneling value $\tau$, leading to tunneling noise. However, the magnitude of tunneling noise is much weaker than that of the detuning noise \cite{koski.20}. Here, the error in the AC-driven field $\varepsilon_{\rm{AC}}$ is neglected, which is also much weaker than the detuning noise \cite{abadillo.19}. In addition, the fluctuation in the Overhauser field (nuclear noise) in GaAs can lead to unwanted error in $\Delta B$ and also introduce relaxation \cite{Reilly.08}. To deal with the nuclear noise, the silicon-based semiconductor quantum dot with isotopic purification is appreciated, where the nuclear noise is strongly suppressed \cite{Huang.19}. Meanwhile, the valley in Si/SiGe platforms may introduce extra valley-spin coupling leading to relaxation \cite{ZhangX.19}. Fortunately, recent experiment indicated that by using the silicon metal-oxide-semiconductor (Si-MOS) platform the valley splitting can be as high as 0.8 meV \cite{Kawakami.14,Marko.16}. In this way, the valley effect can also be safely neglected. According to the latest experiment, the relaxation time based on silicon platform has reached $T_{1}=9\ \rm{s}$ \cite{Ciriano.21}. In this work, we focus on the suppression on $\delta\omega_{q}$ due to the detuning noise, and we simulate the single-qubit gate fidelity by using the two-level structure as shown in Eq.~\ref{eq:hac2}.

\begin{figure}
\includegraphics[width=1.0\columnwidth]{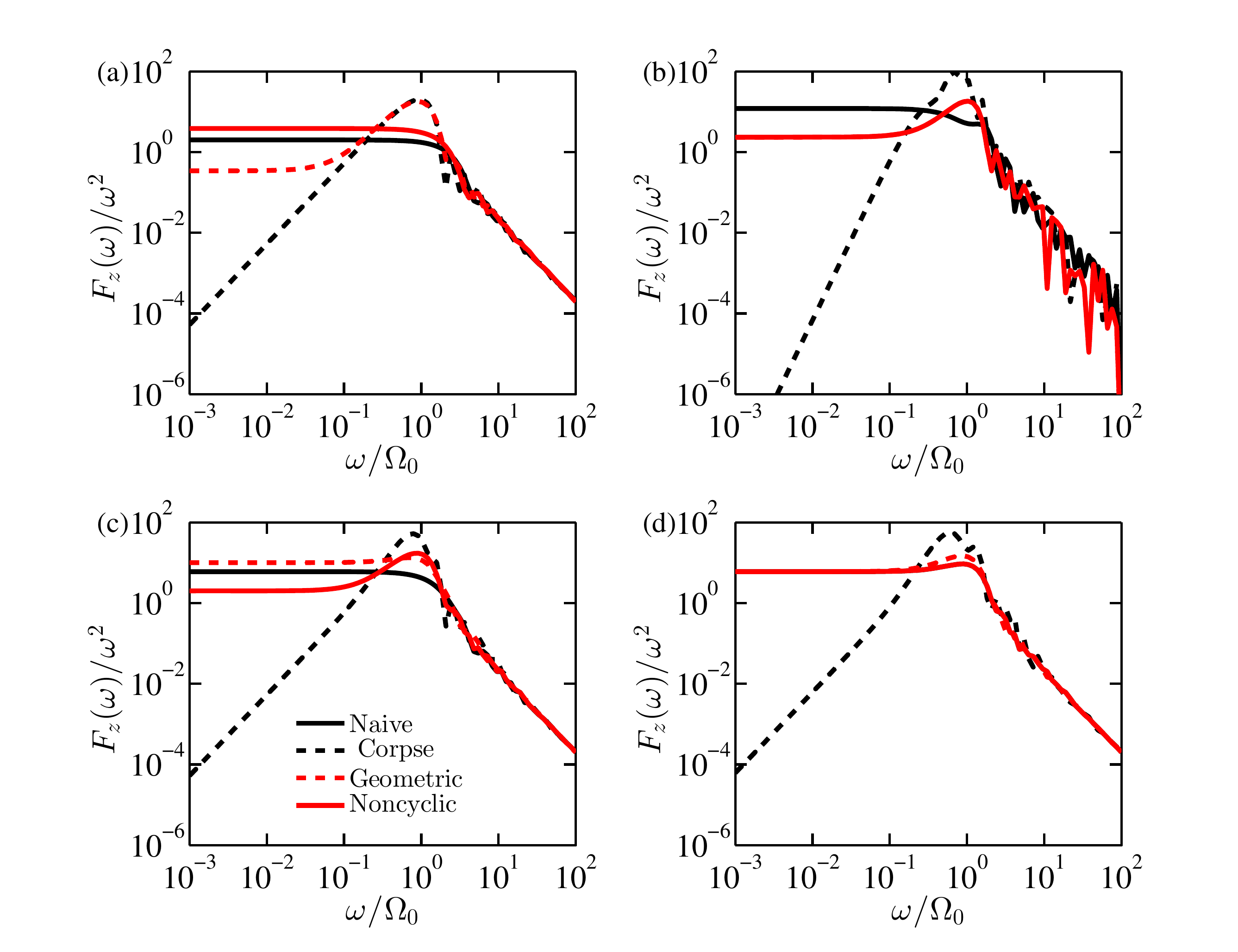}
\caption{Filter transfer function used to calculate individual gate fidelity. The gate in each panel is (a)$R(\hat{x},\pi/2)$, (b)$R(\hat{z},\pi/2)$, (c)$R(\hat{x}+\hat{y}-\hat{z},4\pi/3)$ and (d)$R(\hat{x}+\hat{z},\pi)$. The design of these gates can be seen in Table.~\ref{ap:table}. The fidelity for the naive, the CORPSE, the geometric and the quantum gates are: (a) $\mathcal{F}_{\rm{nai}}=99.78\%$, $\mathcal{F}_{\rm{cor}}=99.61\%$, $\mathcal{F}_{\rm{geo}}=99.59\%$, $\mathcal{F}_{\rm{non}}=99.60\%$. (b) $\mathcal{F}_{\rm{nai}}=98.86\%$, $\mathcal{F}_{\rm{cor}}=98.71\%$, $\mathcal{F}_{\rm{geo}}=99.47\%$, $\mathcal{F}_{\rm{non}}=99.47\%$. (c) $\mathcal{F}_{\rm{nai}}=99.39\%$, $\mathcal{F}_{\rm{cor}}=99.16\%$, $\mathcal{F}_{\rm{geo}}=98.89\%$, $\mathcal{F}_{\rm{non}}=99.47\%$. (d) $\mathcal{F}_{\rm{nai}}=99.29\%$, $\mathcal{F}_{\rm{cor}}=99.01\%$, $\mathcal{F}_{\rm{geo}}=99.18\%$, $\mathcal{F}_{\rm{non}}=99.29\%$. Other parameters: $\Omega_{0}/2\pi=0.045055\ \rm{GHz}$, $S(\omega)=A/(\omega t_{0})^{\alpha}$ with $\alpha=1$ and $At_{0}=10^{-3}$. The cutoffs are $\omega_{\mathrm{ir}}/2\pi=100\ \mathrm{kHz}$ and  $\omega_{\mathrm{uv}}/2\pi=20\ \mathrm{GHz}$.}
\label{fig:filterf}
\end{figure}

In the real evolution process, the noise is time-dependent and is always described by the so-called  $1/f^{\alpha}$ noise model. The power spectral density of the $1/f^{\alpha}$-type noise has the form as $S(\omega)=A/(\omega t_{0})^{\alpha}$ \cite{yang.16}. Here, $A$ denotes the noise amplitude, $t_{0}$ the time unit, and the exponent $\alpha$ the noise correlation. For the semiconductor quantum-dot environment, the charge noise is typically about $\alpha\simeq1$. The noise amplitude can be determined by \cite{zhang.17}
\begin{equation}
\begin{aligned}
\int_{\omega_{\mathrm{ir}}}^{\omega_{\mathrm{uv}}} \frac{A_{\varepsilon}}{\left(\omega t_{0}\right)^{\alpha}} d \omega=\pi\sigma_{\omega q}^{2}.
\label{eq:integ}
\end{aligned}
\end{equation}
Here, we take the cutoffs as $\omega_{\mathrm{ir}}/2\pi=100\ \mathrm{kHz}$ and  $\omega_{\mathrm{uv}}/2\pi=20\ \mathrm{GHz}$ \cite{abadillo.19}. In our simulation, we take the time unit as $t_{0}=1/\Omega_{0}$. Therefore, by substituting $\alpha=1$, we estimate $A_{\varepsilon}t_{0}\simeq10^{-3}$. Here, the $1/f$ noise in the simulation is generated by using the method as described in \cite{yang.16}. On the other hand, for the piecewise continuous Hamiltonian (see Eq.~\ref{eq:hac}), the time-dependent noise spectrum and the infidelity can be well characterized via the filter transfer function \cite{Green.12,Todd.13,Silva.14}. The detail on how to analyze the fidelity using this method can be seen in Ref.~\cite{Todd.13}. Here, we briefly introduce this method and focus on the $z$-component noise. For the desired gate rotation $U(t)$ which satisfies the Schrodinger equation $i \dot{U}(t)=H(t) U(t)$, one can define a specific control matrix \cite{Todd.13}
\begin{equation}
\begin{aligned}
R_{i j}(t)=[\boldsymbol{R}(t)]_{i j}=\operatorname{Tr}[U^{\dagger}(t) \sigma_{i} U(t) \sigma_{j}]/2,
\label{eq:controlt}
\end{aligned}
\end{equation}
with its Fourier transform as
\begin{equation}
\begin{aligned}
R_{i j}(\omega)=-i \omega \int_{0}^{T} d t R_{i j}(t) e^{i \omega t},
\label{eq:controlw}
\end{aligned}
\end{equation}
where $i, j \in\{x, y, z\}$. And then, the average gate fidelity in the noise environment that is determined by the specific cross-spectral density matrix $S_{ij}(\omega)$ can be \cite{Todd.13}
\begin{equation}
\begin{aligned}
\mathcal{F}_{\mathrm{av}} \simeq& 1-\frac{1}{2 \pi} \sum_{i, j, k=x, y, z} \int_{-\infty}^{\infty} \frac{\mathrm{d} \omega}{\omega^{2}} S_{i j}(\omega) R_{j k}(\omega) R_{i k}^{*}(\omega).
\label{eq:fidelity}
\end{aligned}
\end{equation}
Since we only consider the $z$-component noise, the cross-spectral density is with the form of $S_{z}(\omega)$. In this way, Eq.~\ref{eq:fidelity} reduces to 
\begin{equation}
\begin{aligned}
\mathcal{F}_{\mathrm{av}}=&1-\frac{1}{2 \pi} \int_{-\infty}^{\infty} d \omega S_{z}(\omega) \frac{F_{z}(\omega)}{\omega^{2}}
\label{eq:fidelity2}
\end{aligned}
\end{equation}
where
\begin{equation}
\begin{aligned}
F_{z}(\omega)=\sum_{k=x, y, z} R_{z k}(\omega) R_{z k}^{*}(\omega)
\label{eq:filterfun}
\end{aligned}
\end{equation}
is the filter transfer function for the $z$-component detuning noise. In Appendix.~\ref{appx:filter}, we have given a step-by-step example to calculate the filter transfer function. One can then use it to calculate the fidelity.

Considering the $1/f$ noise spectrum, one can further derive the fidelity as
\begin{equation}
\begin{aligned}
\mathcal{F}_{\mathrm{av}}=1- \frac{1}{2 \pi}\int_{\omega_{\mathrm{ir}}}^{\omega_{\mathrm{uv}}} \frac{A_{\epsilon}}{\omega t_{0}} \frac{F_{z}(\omega)}{\omega^{2}}
\label{eq:filter1f}
\end{aligned}
\end{equation}
Note that, to simplify the calculation one can use the unit of $\Omega_{0}$ and the corresponding cutoffs turn to be $\omega_{\mathrm{ir}}'=\omega_{\mathrm{ir}}/\Omega_{0}\simeq 2.220\times 10^{-3}$ and $\omega_{\mathrm{uv}}'=\omega_{\mathrm{uv}}/\Omega_{0}\simeq 4.439\times 10^{2}$.

From Eq.~\ref{eq:fidelity2} it is clear that for a given noise spectrum the related filter transfer function i.e. $F_{z}(\omega)/\omega^{2}$ has positive relationship with the infidelity. Therefore, it can be used to characterize the noise effect. Fig.~\ref{fig:filterf} shows the results of the filter transfer function for the naive, the CORPSE, the geometric and the non-cyclic geometric quantum gates. In the plot, we show four different kinds of gates ($R(\hat{x},\pi/2)$, $R(\hat{z},\pi/2)$, $R(\hat{x}+\hat{y}-\hat{z},4\pi/3)$ and $R(\hat{x}+\hat{z},\pi)$), which are the representatives for the rotation around different axis in the single-qubit Clifford group. The design of these gates can be seen in Table.~\ref{ap:table}. The performance of other gates in the Clifford group is similar and is thus not shown. We find that in the low-frequency region i.e. $\omega/\Omega_{0}<10^{-1}$, the filter transfer function for the CORPSE gate is normally lower than others. This means that the CORPSE gate is more robust to the low-frequency noise.  However, it stands out in the high-frequency region $10^{-1}<\omega/\Omega_{0}<10^{0}$. When the frequency is high enough, the performance for all the gates are not distinguishable. Using these filter transfer function results, we further calculate the gate fidelity as shown in the plot. It is shown that whether the non-cyclic geometric gate can offer improvement over the naive gates depends on the specific noise spectrum up to the gates. For example, for the gate $R(\hat{x},\pi/2)$ in (a), it performs worse than the naive gate. Meanwhile, for the gate $R(\hat{x}+\hat{z},\pi)$ in (d), it performs the same as the naive gate. On the other hand, both $R(\hat{z},\pi/2)$ in (b) and $R(\hat{x}+\hat{y}-\hat{z},4\pi/3)$ in (c) can outperform the naive gate. On the other hand no improvement is offered by the CORPSE gate.

\subsection{Two-qubit gate}\label{sec:twoqubit}

The two-qubit entangling  gate can be implemented by coupling two ST qubits via a cavity resonator as shown in Fig.~\ref{fig:model}(c). In the DQD eigenbasis, the total Hamiltonian of the hybrid system composed of the qubits and the resonator reads \cite{abadillo.19}
\begin{equation}
\begin{aligned}
H_{\rm{tot}}=& \omega_{r} a^{\dagger} a+\sum_{k=1}^{2} \sum_{m=1}^{3} E_{m}^{(k)} \sigma_{mm}^{(k)} \\
	&+\sum_{k=1}^{2} \sum_{m,n=1}^{3} g^{(k)} d_{mn}^{(k)}\left(a+a^{\dagger}\right) \sigma_{mn}^{(k)},
\end{aligned}
\label{fig:Htot}
\end{equation}
where $a^{\dagger}$ ($a$) is the photon creation (annihilation) operator for the resonator and $\omega_{r}$ is the intrinsic frequency of the resonator, while $g^{(k)}$ is the coupling strength between the $k$th DQD and the resonator. $E_{m}^{(k)}$ and $\sigma_{mn}^{(k)}$ are the eigenvalue and the Pauli matrix for the $k$th DQD. 
By moving to the rotating frame defined by $U'=e^{-i H_{0} t}$ where
\begin{equation}
\begin{aligned}
H_{0}=\omega_{r} a^{\dagger} a+\sum_{k=1}^{2} \sum_{m=1}^{3} E_{m}^{(k)} \sigma_{mm}^{(k)},
\end{aligned}
\label{fig:H0}
\end{equation}
one finds
\begin{widetext} 
\begin{equation}
\begin{aligned}
H_{\rm{int}}'=& \sum_{k=1}^{2} \sum_{m=1}^{3} g^{(k)}d_{mm}^{(k)}e^{i \omega_{r} t}\sigma_{mm}^{(k)}a^{\dagger}+
\sum_{k=1}^{2} \sum_{m<n}g^{(k)}d_{mn}^{(k)}(\sigma_{mn}a^{\dagger}e^{i (E_{m}^{(k)}-E_{n}^{(k)}+\omega_{r}) t}+\sigma_{mn}ae^{i (E_{m}^{(k)}-E_{n}^{(k)}-\omega_{r})t})+h.c.
\end{aligned}
\label{Hint2}
\end{equation}
\end{widetext}
In the experiments, the resonator frequency can be as high as several GHz, while the coupling strength is typically less than 100 MHz \cite{abadillo.19}, indicating $|g^{(k)}d_{mn}^{(k)}|/2\pi\leq50$ MHz and therefore we have $|g^{(k)}d_{mm}^{(k)}|\ll\omega_{r}$. Also, we can set the proper parameters in the Hamiltonian Eq.~\ref{eq:hst} to ensure $|g^{(k)}d_{mn}^{(k)}|\ll|E_{m}^{(k)}-E_{n}^{(k)}-\omega_{r}|$. In this way, the leakage to the excited states $|f\rangle$ (related to the terms $d_{gf}$ and $d_{ef}$) and the longitudinal coupling between the qubits and the resonator (related to the terms $d_{gg}$, $d_{ee}$ and $d_{ff}$) are strongly suppressed. This is owing to the fact that they can be regarded as the fast-oscillating terms. Further, if we set $\omega_{q}^{(k)}=\omega_{r}$, namely, both of the qubits are in resonance with the resonator, Eq.~\ref{Hint2} reduces to
\begin{equation}
\begin{aligned}
H_{\rm{int}}'\approx& \sum_{k=1}^{2} \Omega^{(k)}\sigma_{ge}^{(k)}a^{\dagger}+h.c.
\end{aligned}
\label{Hint3}
\end{equation}
where $\Omega^{(k)}=g^{(k)}d_{ge}^{(k)}$. In the single-excitation subspace spanned by $\{|eg0\rangle,|ge0\rangle,|gg1\rangle\}$, Eq.~\ref{Hint3} can establish an effective resonant three-level $\Lambda$ system. Here, $|abc\rangle$ represents the first and the second qubit and the resonator.  When the evolution time satisfies $\int_{0}^{t} \Omega(t) \mathrm{d} t^{\prime}=\pi$, a two-qubit entangling gate is obtained \cite{zhouj.17,egger.19,li.20,zhang.20b}. In the computational subspace, which corresponds to the zero-photon subspace $\{|gg0\rangle,|ge0\rangle,|eg0\rangle,|ee0\rangle\}$ one finds
\begin{equation}
\begin{aligned}
U_{\rm {ent }}\left(\xi\right)=\left(\begin{array}{cccc}
	1 & 0 & 0 & 0 \\
	0 & \cos \xi & \sin \xi & 0 \\
	0 & \sin \xi & -\cos \xi & 0 \\
	0 & 0 & 0 & -1
\end{array}\right),
\end{aligned}
\label{fig:Hgeo2}
\end{equation}
where we have assumed $\Omega=\sqrt{\left(\Omega^{(1)}\right)^{2}+\left(\Omega^{(2)}\right)^{2}}$ and $\tan\xi/2=-\Omega^{(1)}/\Omega^{(2)}$. 

The performance of the two-qubit gate can be evaluated via numerically solving the master equation \cite{blais.07}
\begin{equation}
\begin{aligned}
\dot{\rho}=-i\left[H_{\rm{int}}', \rho\right]+\Gamma_{a} \mathcal{D}[a]+\Gamma_{2} \mathcal{D}[\sigma_{z}] +\Gamma_{1} \mathcal{D}[\sigma],
\end{aligned}
\label{fig:master}
\end{equation}
where 
\begin{equation}
\begin{aligned}
\mathcal{D}[L]= \left(2 L \rho L^{\dagger}-L^{\dagger} L \rho-\rho L^{\dagger} L\right) / 2.
\end{aligned}
\label{fig:master2}
\end{equation}
Here, $\Gamma_{a}$ denotes the internal decay effect of the resonator, while $\Gamma_{1}$ and $\Gamma_{2}$ denote the relaxation and pure dephasing rate, respectively. Meanwhile, we assume the pure dephasing is mainly owing to the charge noise and the fast-oscillating terms, both of which have been considered in the Hamiltonian $H_{\rm{int}}'$, we therefore set $\Gamma_{2}=0$. Note that, the relaxation of the ST qubits is complicated. However, as stated above, the relaxation rate in the silicon-based platform can be substantially low, since the relaxation time in the experiment has reached $T_{1}=9\ \rm{s}$ \cite{Ciriano.21}. Therefore, the relaxation of the coupled system can be mainly owing to the leakage to the excited state $|f\rangle$ and the resonator decay.

\begin{figure}
	\includegraphics[width=1.0\columnwidth]{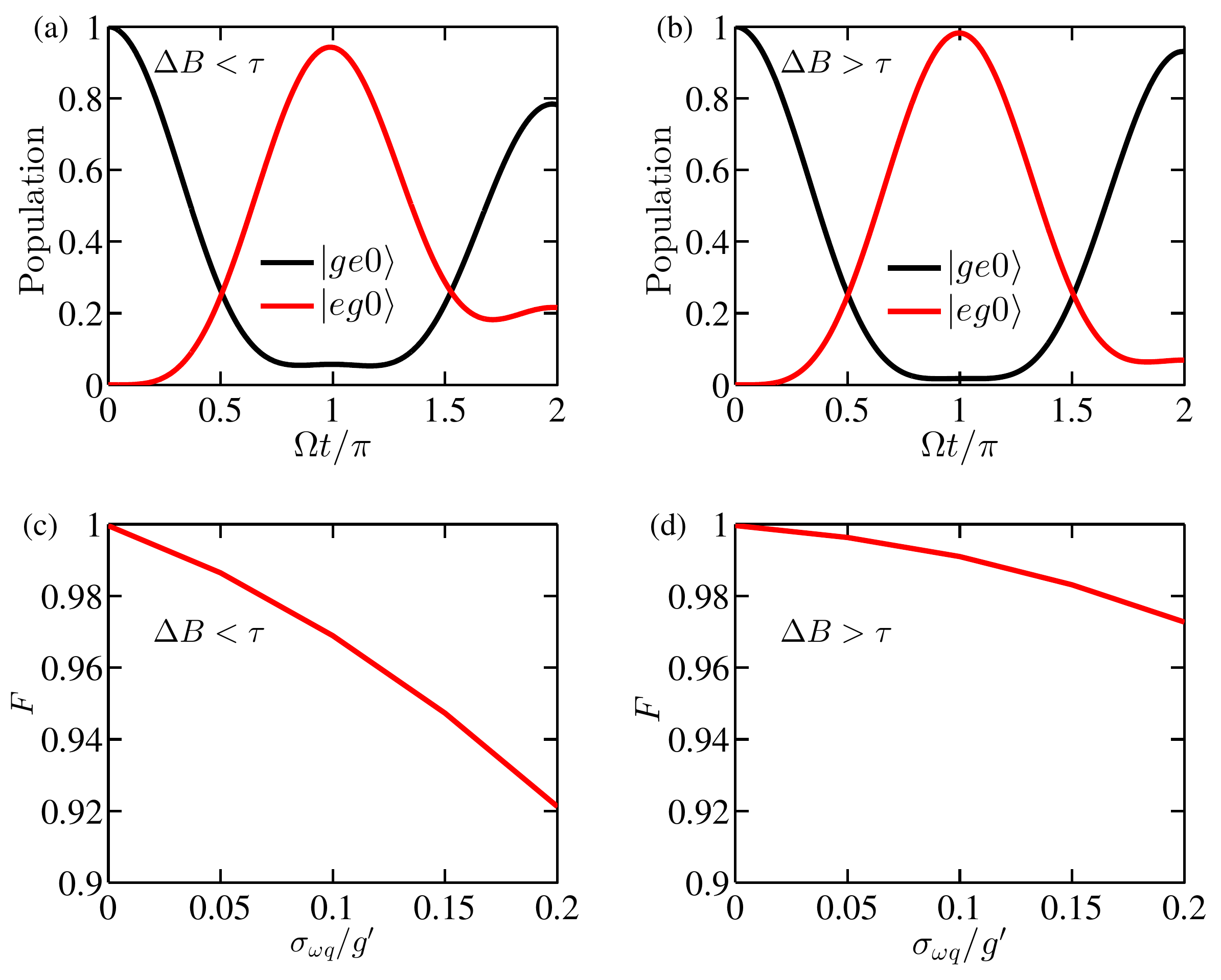}
	\caption{(a) and (b): the population of the states $|ge0\rangle$ and $|eg0\rangle$. (c) and (d): the fidelity of the  entangling gate $U_{\rm {ent }}\left(\xi=-\pi/2\right)$ as a function of $\sigma_{\omega q}/g'$. Parameters: for (a) and (c): $\Delta B/2\pi=1.5\ \rm{GHz}$, $\tau/2\pi=1.75\ \rm{GHz}$ while for  (b) and (d): $\Delta B/2\pi=2.5\ \rm{GHz}$, $\tau/2\pi=1.5\ \rm{GHz}$. The coupling strength is taken as $g^{(1)}=g^{(2)}=g'=2\pi\times 100\ \rm{MHz}$. The resonator decay is taken as $\Gamma_{a}/2\pi=0.028\ \rm{MHz}$, which corresponds to the resonator quality factor of about $Q=10^{5}$ \cite{samkharadze.16}.}
	\label{fig:twoqubit}
\end{figure}

Under the action of the entangling gate $U_{\rm {ent }}\left(\xi=-\pi/2\right)$ which corresponds to $g^{(1)}=g^{(2)}=g'$, the initial state $|ge0\rangle$ will transform into state $|eg0\rangle$ at the final evolution time. In Fig.~\ref{fig:twoqubit}(a) and (b) we show the population of the states $|ge0\rangle$ and $|eg0\rangle$, which corresponds to the cases $\Delta B<\tau$ and $\Delta B>\tau$, respectively. Here, we set $g'=2\pi\times 100\ \rm{MHz}$. The resonator decay is taken as $\Gamma_{a}/2\pi=0.028\ \rm{MHz}$ according to the recent experiment, which corresponds to the resonator quality factor of about $Q=10^{5}$ \cite{samkharadze.16}. The population of $|eg0\rangle$ in Fig.~\ref{fig:twoqubit}(a) is obviously smaller than the case in Fig.~\ref{fig:twoqubit}(b) due to the strong leakage effect. To deeply study the robustness of the  two-qubit gate, we further plot the fidelity as a function of $\sigma_{\omega q}/g'$, as shown in Fig.~\ref{fig:twoqubit} (c) and (d). To ensure convergence, we have averaged the fidelities over 1000 implementations for each $\sigma_{\omega q}/g'$. When the noise is small, i.e. $\sigma_{\omega q}/g'\simeq0$, the fidelities for both cases are closed to 1. As $\sigma_{\omega q}/g'$ increases, the fidelity in Fig.~\ref{fig:twoqubit}(c) drops quickly, and it is down to 92\% for the large noise of $\sigma_{\omega q}/g'\simeq0.2$. For comparison, the fidelity in Fig.~\ref{fig:twoqubit}(d) falls more slowly. When $\sigma_{\omega q}/g'\simeq0.2$, the fidelity is still surpassing 97\%. Thus, to achieve high-fidelity entangling gate, one is wise to operate it in the region $\Delta B>\tau$, which is consistent with the single-qubit case. Assuming that the charge noise here is with the same level as that of the single-qubit case, namely $\sigma_{\omega q}/g'\simeq0.05$, the fidelity in Fig.~\ref{fig:twoqubit}(d) can be as high as 99.63\%.

\section{Conclusions }\label{sec:conclusion}

We theoretically propose the universal gates for the ST qubits. We have shown that the single-qubit gates can be achieved by introducing an AC drive on the detuning near the TSS. The large dipole moments of the DQDs at the TSS has enabled strong coupling between the qubits and the cavity resonator, which leads to a two-qubit entangling gates. We have discussed the implementation of the gates in two different regions for the TSS. When operating in the region $\varepsilon_{\rm{SS}}>0$, both the single- and two-qubit gates are having fidelity higher than 99\%. Our method offers an alternative tool to realize high-fidelity ST qubits.

\section*{ACKNOWLEDGMENTS}\label{sec:ack}
This work was supported by Key-Area Research and Development Program of GuangDong Province  (Grant No. 2018B030326001), the National Natural Science Foundation of China (Grant No. 11905065, 11874156), the Project funded by China Postdoctoral Science Foundation (Grant No. 2019M652928), the Science and Technology Program
of Guangzhou (Grant No. 2019050001).

\appendix



\section{Effective two-level Hamiltonian and the RWA effect}\label{appx:Heff}
\begin{figure}
	\includegraphics[width=0.8\columnwidth]{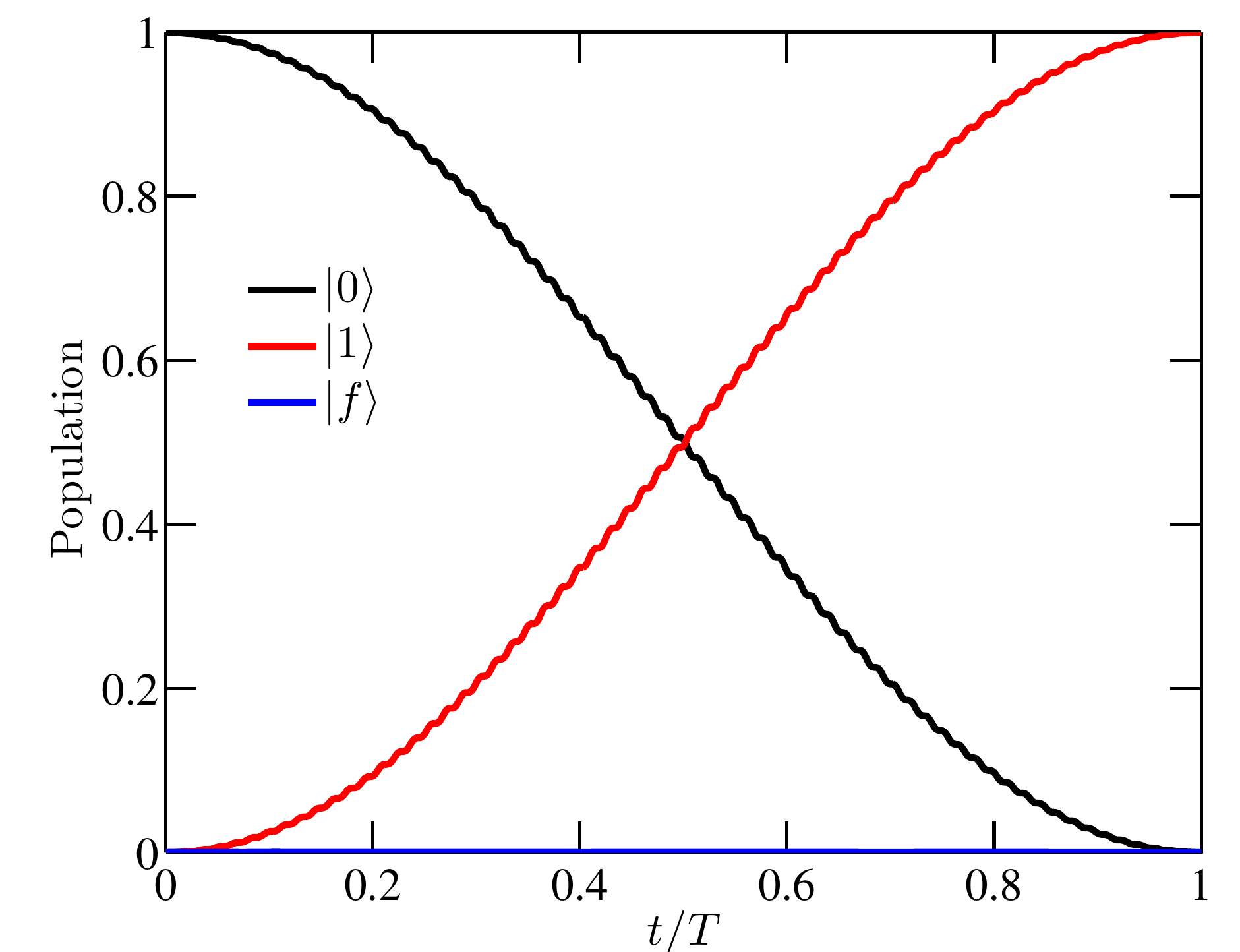}
	\caption{The population of the computational states $|0\rangle$ and $|1\rangle$, and the leakage state $|f\rangle$ when operating a $\pi$-pulse rotation. Parameters: $\Delta B/2\pi=2.5\ \rm{GHz}$, $\tau/2\pi=1.5\ \rm{GHz}$ and $\varepsilon_{\rm{AC}}/2\pi=0.1\ \rm{GHz}$. For a given initial state $|0\rangle$, the population of the state $|f\rangle$ (denoted as black line) is always with the order of $10^{-4}$. On the other hand, the population of state $|0\rangle$ at the final evolution time is with the order of $10^{-6}$.}
	\label{fig:app}
\end{figure}

Here, we seek how to obtain the effective two-level Hamiltonian for the single-qubit quantum gate and estimate the RWA effect. The total Hamiltonian written in the eigenstates considering the detuning noise reads
\begin{equation}
H=\sum_{n=1}^{3} (E_{n}(\varepsilon_{\rm{SS}})+\delta E_{n}) \sigma_{n n}+\varepsilon' \sum_{m, n=1}^{3} d_{mn} \sigma_{m n}
\label{ap:h},
\end{equation}
where $E_{n}$ is the $n$th energy level (the number 1, 2, and 3 here represent states $|g\rangle$, $|e\rangle$ and $|f\rangle$), $\delta E_{n}$ is the drift of the energy level induced by the detuning noise, and $\varepsilon'(t)=\varepsilon_{\rm{AC}}\cos(\omega t+\phi(t))$. In the rotating frame defined by $U_i=e^{-i (\sum_{n=1}^{3}E_{n} |n\rangle\langle n|)t}$, we further have
\begin{equation}
\begin{aligned}
H_{\rm{rot}} &= U_{i}^{\dagger}H U_{i}-iU_{i}^{\dagger}\frac{\partial U_{i}}{\partial t}
\\
=&H_{\rm{leak}}+H_{\rm{compu}},
\end{aligned}
\label{eq:Hi}
\end{equation}
where
\begin{widetext}
\begin{equation}
\begin{aligned}
\begin{small}
H_{\rm{leak}}=\frac{\varepsilon_{\rm{AC}}}{2}
\left(\begin{array}{ccc}
d_{ff}(e^{-i(\phi+\omega t)}+e^{i(\phi+\omega t)})+\delta E_{f}& d_{ef}(e^{-i(\phi+\omega t-\omega_{ef}t)}+e^{i(\phi+\omega t+\omega_{ef}t)}) &d_{gf}(e^{-i(\phi+\omega t-\omega_{gf}t)}+e^{i(\phi+\omega t+\omega_{gf}t)})   \\
d_{ef}(e^{i(\phi+\omega t-\omega_{ef}t)}+e^{-i(\phi+\omega t+\omega_{ef}t)})   &0& 0\\
d_{gf}(e^{i(\phi+\omega t-\omega_{gf}t)}+e^{-i(\phi+\omega t+\omega_{gf}t)})  &0& 0		
\end{array}\right),
\end{small}
\end{aligned}
\end{equation}
\end{widetext}
and
\begin{widetext}
\begin{equation}
\begin{aligned}
\begin{small}
H_{\rm{compu}}=	\left(\begin{array}{ccc}
0   &      0    &   0   \\
0   &      	\frac{\varepsilon_{\rm{AC}}d_{ee}}{2}(e^{-i(\phi+\omega t)}+e^{i(\phi+\omega t)})+\delta E_{e}    &   \frac{\varepsilon_{\rm{AC}}d_{ge}}{2}(e^{-i(\phi+\omega t-\omega_{q}t)}+e^{i(\phi+\omega t+\omega_{q}t)})\\
0   &      \frac{\varepsilon_{\rm{AC}}d_{ge}}{2}(e^{i(\phi+\omega t-\omega_{q}t)}+e^{-i(\phi+\omega t+\omega_{q}t)})    &   	\frac{\varepsilon_{\rm{AC}}d_{gg}}{2}(e^{-i(\phi+\omega t)}+e^{i(\phi+\omega t)})+\delta E_{g}		
\end{array}\right),
\end{small}
\end{aligned}
\end{equation}
\end{widetext}
Here we can see that the counter-rotating terms appear in both $H_{\rm{leak}}$ and $H_{\rm{compu}}$, the former of which leads to leakage to the excited state $|f\rangle$, while the latter would result in gate error. Consider the RWA condition is met, namely, $|\frac{\varepsilon_{\rm{AC}}d_{mn}}{2}|\ll \omega,|\omega\pm\omega_{gf}|,|\omega\pm\omega_{ef}|$, and the resonance condition is also satisfied, i.e. $\omega_{q}=\omega$, then, all the counter-rotating terms vanish. The Hamiltonian is reduced to the effective two-level structure as
\begin{equation}
\begin{aligned}
H_{\rm{compu}}^{'}=	\left(\begin{array}{cc}
\frac{\delta \omega_{q}} {2}   &   \frac{\varepsilon_{\rm{AC}}d_{ge}}{2}e^{-i\phi}\\
\frac{\varepsilon_{\rm{AC}}d_{ge}}{2}e^{i\phi}    &   -\frac{\delta \omega_{q}} {2}		
\end{array}\right)\\
=\frac{\Omega_{0}}{2}\left(\cos \phi\ \sigma_{x}+\sin \phi\ \sigma_{y}\right)+ \frac{\delta}{2}\ \sigma_{z}
\end{aligned}
\label{ap:heff2}
\end{equation}
where $\Omega_{0}=|d_{ge}|\varepsilon_{\rm{AC}}$ and $\delta=\delta\omega_{q}=\delta E_{e}-\delta E_{g}$. To suppress the counter-rotating terms, one should carefully choose the parameters for the TSS point and the value of $\varepsilon_{\rm{AC}}$ should not be too large. For our chosen parameters as shown in the main text, Eq.~\ref{ap:heff2} is always valid. This can be estimated by plotting the evolution of all the three eigenstates when operating a $\pi$-pulse rotation (i.e. a NOT gate). The evolution of the states can be numerically calculated via solving the von Neumann equation
\begin{equation}
\begin{aligned}
i \hbar \frac{\partial \rho}{\partial t}=[H_{\rm{rot}}, \rho]
\end{aligned}
\label{ap:rou}
\end{equation}
As shown in Fig.~\ref{fig:app}, for a given initial state $|0\rangle$ (denoted as black line), the population of the state $|f\rangle$ (denoted as blue line) is always with the order of $10^{-4}$. This means that the leakage to the excited state $|f\rangle$ is strongly suppressed and thus can be safely ignored. On the other hand, the population of state $|0\rangle$ at the final evolution time is with the order of $10^{-6}$. This indicates that the gate error due to the counter-rotating terms in the computational subspace is substantially weak. Therefore, the RWA is highly effective, and we would focus on the two-level Hamiltonian as described in Eq.~\ref{ap:heff2}

\section{Expansion of the fidelity}\label{appx:fidelity}

Here, we analytically compare the fidelity of the naive, the CORPSE, the geometric and the quantum gates. We first write down the evolution operators for each type of the gates and then derive the fidelity expressions. 

\begin{figure}
	\includegraphics[width=1\columnwidth]{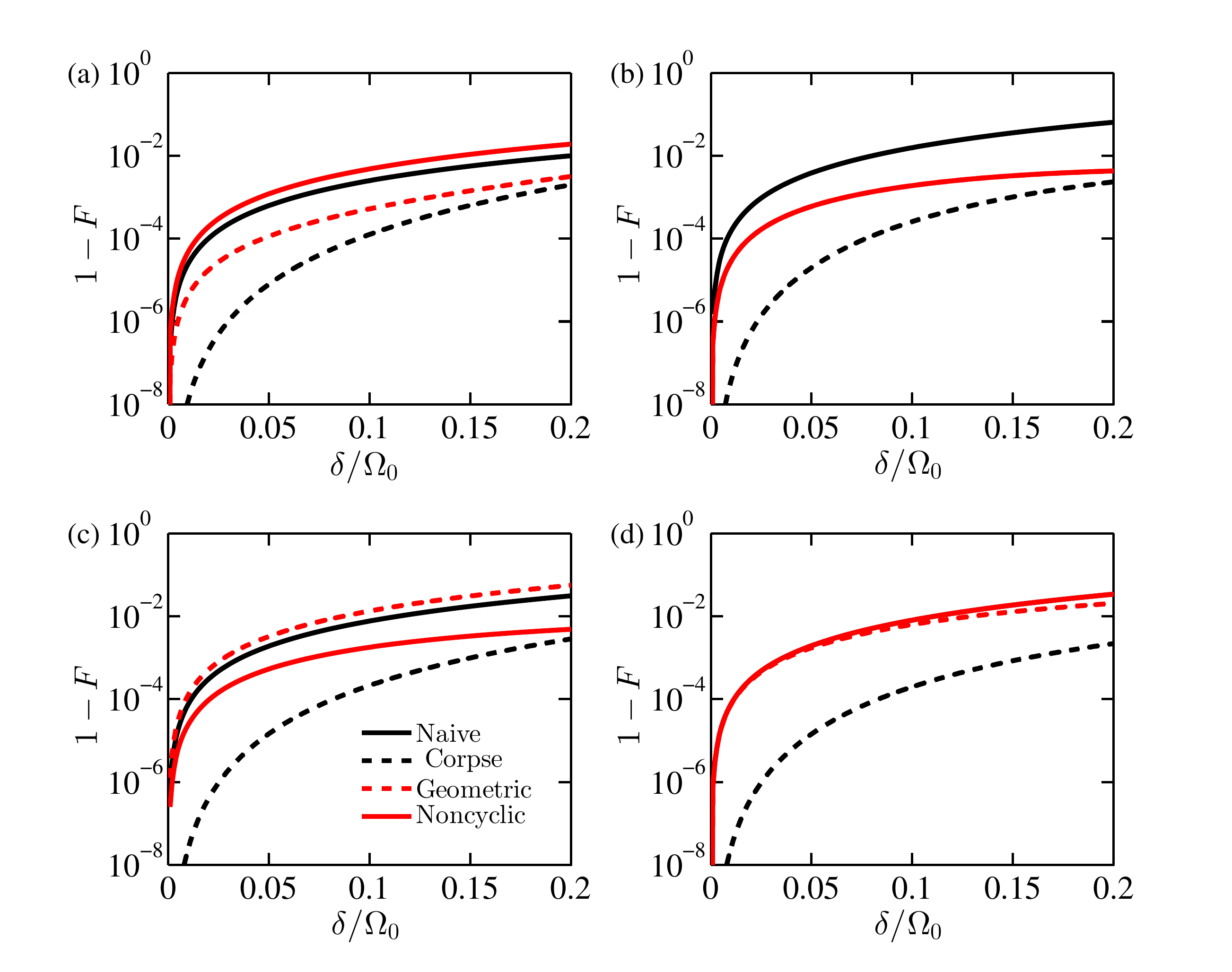}
\caption{Infidelity for several gates. The gate in each panel is (a)$R(\hat{x},\pi/2)$, (b)$R(\hat{z},\pi/2)$, (c)$R(\hat{x}+\hat{y}-\hat{z},4\pi/3)$ and (d)$R(\hat{x}+\hat{z},\pi)$. The design of these gates can be seen in Table.~\ref{ap:table}.}
\label{fig:app2}
\end{figure}

In Sec.~\ref{appx:Heff}, we have shown that the system can be treated well as a two-level system, where the dominant detuning noise appears in the diagonal term. Therefore, we would calculate the fidelity by using the expression in Eq.~\ref{ap:heff2}. For this piecewise-continuous Hamiltonian, the corresponding one-piece evolution operator suffering noise reads
\begin{equation}
\begin{aligned}
R_{\delta}(\phi, \theta) \equiv \exp \left[-i   (\frac{\Omega_{0}}{2}\left(\cos \phi\ \sigma_{x}+\sin \phi\ \sigma_{y}\right)+ \frac{\delta}{2}\ \sigma_{z}) \frac{\theta}{\Omega_{0}} \right]
\end{aligned}
\label{ap:rotation}
\end{equation}
Here, $R_{\delta}(\phi, \theta)$ denotes arbitrary rotation around the axis in the $x$-$y$ plane by an angle $\theta$, where the axis on the Bloch sphere is determined by $\boldsymbol{r}=(\cos \phi,\sin \phi, 0)$. In our case, the single-piece evolution operator $R_{\delta}(\phi, \theta)$ is regarded as the so-called naive gate.

Typically, $R_{\delta}(\phi, \theta)$ can be dynamically corrected by using the CORPSE gates which includes 3 elementary pulse $R_{\delta}(\phi, \theta)$ \cite{Cummins.03,Bando.13}: 
\begin{equation}
\begin{aligned}
R_{\rm{corpse}}(\phi, \theta)=R_{\delta}(\phi_{c3}, \theta_{c3}).R_{\delta}(\phi_{c2}, \theta_{c2}).R_{\delta}(\phi_{c1}, \theta_{c1}).
\end{aligned}
\label{ap:corp}
\end{equation}
where 
\begin{equation}
\begin{aligned}
\begin{array}{l}
	\theta_{c1}=2n_{1}\pi+\theta / 2-\arcsin [\sin (\theta / 2) / 2], \\
	\theta_{c2}=2n_{2}\pi-2 \arcsin [\sin (\theta / 2) / 2], \\
	\theta_{c3}=2n_{3}\pi+\theta / 2-\arcsin [\sin (\theta / 2) / 2], \\
	\phi_{c1}=\phi_{c2}-\pi=\phi_{c3}=\phi.
\end{array}
\end{aligned}
\label{ap:corpse2}
\end{equation}
where $n_{i}$ ($i=1,2,3$) is being integer. Here, we consider using the short CORPSE pulse, which corresponds to $n_{1}=n_{3}=0$, $n_{2}=1$.

For our quantum gate as shown in Eq.~\ref{eq:noncy} in the main text, the corresponding evolution operator is 
\begin{equation}
\begin{aligned}
R_{\rm{non}}(\chi _0,\phi _{0},\phi _{\rm{1}},\beta _0)=R_{\delta}(\phi_{n2}, \theta_{n2}).R_{\delta}(\phi_{n1}, \theta_{n1})
\end{aligned}
\label{ap:ngeo}
\end{equation}
where 
\begin{equation}
\begin{aligned}
\begin{array}{l}
\theta_{q1}=\chi_{0}, \\
\theta_{q2}=\beta_{0}, \\
\phi_{q1}=\phi_{0}+\pi/2,\\
\phi_{q2}=\phi_{0}+\phi_{1}+\pi/2.
\end{array}
\end{aligned}
\label{ap:ngeo2}
\end{equation}
Using $R_{\rm{non}}(\chi _0,\phi _{0},\phi _{\rm{1}},\beta _0)$ one is able to design arbitrary rotation. For example, the rotation around $x$ axis can be designed as $R_{\rm{non}}(\chi _0,\phi _{0}=\frac{\pi}{2},\phi _{\rm{1}}=\pi,\beta _0)$, and the rotation angle is $\beta_{0}-\chi_{0}$.

On the other hand, the geometric gates considered in this paper are also based on the 3 elementary pulse $R_{\delta}(\phi, \theta)$ \cite{Zhao.17,zhang.20a}:
\begin{equation}
\begin{aligned}
&R_{\rm{geo}}(\theta',\phi',\gamma')=e^{-i \frac{\gamma'}{2} \vec{n} \cdot \vec{\sigma}}\\
=&R_{\delta}(\phi_{\rm{geo}3}, \theta_{\rm{geo}3}).R_{\delta}(\phi_{\rm{geo}2}, \theta_{\rm{geo}2}).R_{\delta}(\phi_{\rm{geo}1}, \theta_{\rm{geo}1}).
\end{aligned}
\label{ap:geo}
\end{equation}
where 
\begin{equation}
\begin{aligned}
\begin{array}{l}
\theta_{\rm{geo}1}=\theta', \\
\theta_{\rm{geo}2}=\pi, \\
\theta_{\rm{geo}3}=\pi-\theta', \\
\phi_{\rm{geo}1}=\phi-\pi/2,\\
\phi_{\rm{geo}2}=\phi-\gamma'/2-\pi/2,\\
\phi_{\rm{geo}3}=\phi-\pi/2.
\end{array}
\end{aligned}
\label{ap:geo2}
\end{equation}
and $\vec{n}=(\sin \theta' \cos \phi', \sin \theta' \sin \phi', \cos \theta')$.

Here, we treat $\delta$ as a quasi-static perturbation with $|\delta|\ll1$, then, we can expand the fidelity for each type of gates by using the evolution operator above. It is well known that any single-qubit gate can be further decomposed into a ``$x$-$y$-$x$'' rotation \cite{nielsen.02}:
\begin{equation}
R_{x}\left(\phi_{a}=0,\theta_{a}\right) R_{y}\left(\phi_{b}=\pi/2,\theta_{b}\right) R_{x}\left(\phi_{c}=0,\theta_{c}\right) ,
\end{equation}
Therefore, we can focus on the performance of the rotation around the $x$ ($y$) axis by an arbitrary rotation angle $\gamma$. The corresponding fidelity expressions around the $x$ axis for each gate are
\begin{equation}
\begin{aligned}
\begin{array}{l}
	\mathcal{F}_{\delta}(\hat{x}, \gamma)\approx\frac{1}{4}\left(4-\delta^{2}+\delta^{2} \cos\gamma\right), \\
	\mathcal{F}_{\rm{non}}(\hat{x}, \gamma)\approx1+\frac{1}{4} \delta^{2}(-3+2 \cos\beta_{0}-\cos\gamma+2 \cos\chi_{0}),\\
	\mathcal{F}_{\rm{geo}}(\hat{x}, \gamma)\approx1-\frac{3 \delta^{2}}{4}+\delta^{2} \cos \frac{\gamma}{2}-\frac{1}{4} \delta^{2} \cos\gamma,
\end{array}
\end{aligned}
\label{ap:fidelity}
\end{equation}
and 
\begin{widetext}
\begin{equation}
\begin{aligned}
&\mathcal{F}_{\rm{corpse}}(\hat{x}, \gamma)\approx\frac{1}{32}\left(-32+\left(7+(-2 \pi+\gamma)^{2}\right) \delta^{4}\right)\\
-&\frac{1}{32}\delta^{4}\left(6 \cos \gamma+\cos2 \gamma+2 \sin \frac{\gamma}{2}\left((4 \pi-2 \gamma) \cos\frac{\gamma}{2}+\sqrt{2} \sqrt{7+\cos\gamma}(-2 \pi+\gamma+\sin\gamma)\right)\right).
\end{aligned}
\label{ap:fidcorpse}
\end{equation}
\end{widetext}
Here, we see that the first order effect of the detuning noise vanishes for all the gates. Meanwhile, CORPSE gate can eliminate the noise effect up to 4th order, which outperforms other gates. We also find that the gate fidelity depends on the rotation angle $\gamma$. The non-cyclic geometric gate can outperform the naive gate when $-\pi<\gamma<\pi$.  On the other hand, its performance still depends on the choice of $\chi_{0}$ and $\beta_{0}$. For a simple choice of $\chi_{0}=0.01 \gamma$ and $\beta_{0}=1.01\gamma$ ($\gamma=\beta_{0}-\chi_{0}$), the non-cyclic geometric gate can correct the naive gate in the region between $-2\pi<\gamma<-\pi$ and $\pi<\gamma<2\pi$. The case for the rotation around the $y$ axis is similar and we would not discuss it in detail. In Fig.~\ref{fig:app2}, we show the infidelity for several considered gates when suffering noise.

\section{An example to calculate filter transfer function}\label{appx:filter}

Here, we introduce how to step-by-step calculate the piecewise continuous filter function. We take a special case of the short CORPSE gate for example, i.e. $R(\hat{x}, \pi/2)$, which includes three piecewise pulse. The case for other types of gate is just a simplified version. For $R(\hat{x}, \pi/2)$ with $\phi=0$ and $\theta=\pi / 2$, we can derive the useful parameters from Eq.\ref{ap:corpse2} as: $\theta_{c 1}=\theta_{c 3}=\pi / 4-\arcsin \left(\frac{1}{2 \sqrt{2}}\right), \theta_{c 2}=2 \pi-2 \arcsin \left(\frac{1}{2 \sqrt{2}}\right)$, $\phi_{1 c}=\phi_{3 c}=0$ and $\phi_{2 c}=\pi$. The corresponding piecewise Hamiltonian are thus $H_{1}=\frac{\Omega_{0}}{2} \sigma_{x}, H_{2}=-\frac{\Omega_{0}}{2} \sigma_{x}, H_{3}=\frac{\Omega_{0}}{2} \sigma_{x}$. The time interval is $T_{k}=\theta_{c k} / \Omega_{0}$ for $k=1,2,3$.

To calculate the piecewise control matrix as shown in Eq.\ref{eq:controlt}, define the evolution operator during each interval as 
$U_{k}(t)=e^{-i  t H_{k}} $. Let $V_{i j}[U]=\operatorname{Tr}\left(U^{\dagger} \sigma_{i} U \sigma_{j}\right) / 2$.
The corresponding piecewise control matrix is therefore $R_{i j}^{(k)}(\omega)=-i \omega \int_{0}^{T_{k}} d t e^{i \omega t} V_{i j}\left[U_{k}(t)\right]$. We first calculate the indefinite integral for the control matrix with $R'^{(k)}(t)=-i \omega \int d t e^{i \omega t} V_{i j}\left[U_{k}(t)\right]$, where
\begin{widetext}
	\begin{equation}
	\begin{aligned}
	R'^{(1)}(t)=R'^{(3)}(t)=
	\left(\begin{array}{ccc}
	-e^{i\omega t}& 0 & 0  \\
	 0           & -\frac{\omega e^{i\omega t}(\omega\cos\Omega_{0}t-i\Omega_{0}\sin\Omega_{0}t)}{\omega^{2}-\Omega_{0}^{2}} & \frac{\omega e^{i\omega t}(i\Omega_{0}\cos\Omega_{0}t+\omega\sin\Omega_{0}t)}{\omega^{2}-\Omega_{0}^{2}}\\
     0           & -\frac{\omega e^{i\omega t}(i\Omega_{0}\cos\Omega_{0}t+\omega\sin\Omega_{0}t)}{\omega^{2}-\Omega_{0}^{2}} & -\frac{\omega e^{i\omega t}(\omega\cos\Omega_{0}t-i\Omega_{0}\sin\Omega_{0}t)}{\omega^{2}-\Omega_{0}^{2}}		
	\end{array}\right),
	\end{aligned}
	\end{equation}
\end{widetext}
and
\begin{widetext}
	\begin{equation}
	\begin{aligned}
	R'^{(2)}(t)=
	\left(\begin{array}{ccc}
	-e^{i\omega t}& 0 & 0  \\
	0           & -\frac{\omega e^{i\omega t}(\omega\cos\Omega_{0}t-i\Omega_{0}\sin\Omega_{0}t)}{\omega^{2}-\Omega_{0}^{2}} & -\frac{\omega e^{i\omega t}(i\Omega_{0}\cos\Omega_{0}t+\omega\sin\Omega_{0}t)}{\omega^{2}-\Omega_{0}^{2}}\\
	0           & \frac{\omega e^{i\omega t}(i\Omega_{0}\cos\Omega_{0}t+\omega\sin\Omega_{0}t)}{\omega^{2}-\Omega_{0}^{2}} & -\frac{\omega e^{i\omega t}(\omega\cos\Omega_{0}t-i\Omega_{0}\sin\Omega_{0}t)}{\omega^{2}-\Omega_{0}^{2}}		
	\end{array}\right),
	\end{aligned}
	\end{equation}
\end{widetext}
Then, one can easily compute the integral results $R^{(k)}(\omega)$. According to Eq.(29) in Ref.\cite{Todd.13}, one can further define a matrix $\mathbf{\Lambda}^{(k)}$ where
\begin{equation}
\mathbf{\Lambda}^{(k)}=V[Q_{k}], \ \ Q_{k}=U_{k}(T_{k})U_{k-1}(T_{k-1}),...,U_{1}(T_{1})
\end{equation}
In this way, the control matrix is 
\begin{equation}
\boldsymbol{R}(\omega)=\sum_{k=1}^{n} e^{i \omega T'_{k-1}} \boldsymbol{R}^{k}(\omega) \boldsymbol{\Lambda}^{(k-1)}
\end{equation}
where $T'_{k-1}=T_{1}+T_{2}+,...,+T_{k-1}$.

Considering the CORPSE case for $R(\hat{x}, \pi/2)$ with $n=3$, we have 
\begin{equation}
\begin{aligned}
\boldsymbol{\Lambda}^{(1)}&=V\left[Q_{1}\right]=V\left[e^{-i\theta_{c 1} \sigma_{x} / 2}\right]\\ \boldsymbol{\Lambda}^{(2)}&=V\left[Q_{2}\right]=V\left[e^{-i\left(-\theta_{c 2}+\theta_{c 1}\right) \sigma_{x} / 2}\right]
\end{aligned}
\end{equation}
and further
\begin{equation}
\boldsymbol{R}(\omega)=\boldsymbol{R}^{(1)}(\omega)+e^{i \omega T_{1}} \boldsymbol{R}^{(2)}(\omega) \boldsymbol{\Lambda}^{(1)}+e^{i \omega\left(T_{2}+T_{1}\right)} \boldsymbol{R}^{(3)}(\omega) \boldsymbol{\Lambda}^{(2)}
\label{app:rw}
\end{equation}
By inserting the $z$-component of matrix $\boldsymbol{R}(\omega)$ into Eq.\ref{eq:filterfun}, one can obtain the analytical filter function as
\begin{widetext}
\begin{equation}
\begin{aligned}
\frac{F_{z}(\omega)}{\omega^{2}}=&\frac{2}{(\omega ^2-\Omega_{0}^2)^{2}}(\omega^2-2^{\frac{6 \omega }{\Omega_{0}}}(\sqrt{7}+i)^{-\frac{4 \omega }{\Omega_{0}}}e^{\frac{4 i \omega  \csc ^{-1}(2 \sqrt{2})}{\Omega_{0}}}\Omega_{0}^{2}
(-5+(\sqrt{7}+1)\cos\frac{\omega(\pi-4\sec^{-1}(2\sqrt{2}))}{4\Omega_{0}}+
\\
&3 \cos \frac{\omega  (\pi +2 \sec ^{-1}(2\sqrt{2}))}{\Omega_{0}}-(\sqrt{7}-1)  \cos\frac{3 \omega  (\pi +4 \sec^{-1}(2\sqrt{2}))}{4 \Omega_{0}})+\\
&2^{\frac{12 \omega }{\Omega_{0}}}(\sqrt{7}+i)^{-\frac{8 \omega }{\Omega_{0}}}e^{\frac{8i\omega\csc^{-1}(2\sqrt{2})}{\Omega_{0}}}\omega\Omega_{0}
((\sqrt{7}-1)\sin\frac{\omega(\pi-4\sec^{-1}(2\sqrt{2}))}{4\Omega_{0}}+\\
&(\sqrt{7}+1)\sin\frac{3\omega(\pi+4\sec^{-1}(2\sqrt{2}))}{4\Omega_{0}}-2\sin\frac{\omega(\pi+8\sec^{-1}(2\sqrt{2}))}{2\Omega_{0}})
);
\end{aligned}
\end{equation}
\end{widetext}

\section{Rotations used for the filter function}\label{appx:table}

As shown in Table.~\ref{ap:table}, the specific gates are used to calculate fidelity using the transfer function. The CORPSE gates are not listed since they are just the simple transform of the naive gates according to Eq.~\ref{ap:corp}.
\begin{table*}
	\caption{Gates used for simulation.}
	\centering
	\scalebox{0.9}{
	\begin{tabularx}{18cm}{lXlXlXl}
		\hline
		\hline
	Element & Naive & Geometric & Non-cyclic geometric \\ 
		\hline
		$R\left(\hat{x}, \frac{\pi}{2}\right)
		$ & $ R\left(\hat{x}, \frac{\pi}{2}\right) $ & $R_{\rm{geo}}\left(\frac{\pi}{2}, 0,-\frac{\pi}{4}\right)
		$ & $ R_{\rm{non}}\left(\frac{\pi}{8}, \frac{\pi}{2}, \pi, \frac{5 \pi}{8}\right)
		$ \\
		$R\left(\hat{z}, \frac{\pi}{2}\right)
		$ & $ R\left(\hat{x}, \frac{\pi}{2}\right) R\left(\hat{y}, \frac{\pi}{2}\right) R\left(-\hat{x}, \frac{\pi}{2}\right)  $ & $ R_{\rm{geo}}\left(0,0,-\frac{\pi}{4}\right)
		$ & $ R_{\rm{non}}\left(\pi, 0, \frac{\pi}{4}, \pi\right)
		$ \\
		$R(\hat{x}+\hat{z}, \pi)$& $ R\left(-\hat{y}, \frac{\pi}{2}\right) R(\hat{x}, \pi)  $ & $ R_{\rm{geo}}\left(\frac{\pi}{4}, 0,-\frac{\pi}{2}\right)
		$ & $ R_{\rm{non}}\left(\frac{\pi}{2}, 0,-\frac{\pi}{2}, \pi\right)
		$ 
		\\
		$R\left(\hat{x}+\hat{y}-\hat{z}, \frac{4 \pi}{3}\right)
		$ & $R\left(-\hat{x}, \frac{\pi}{2}\right) R\left(-\hat{y}, \frac{\pi}{2}\right)
		$ & $ R_{\rm{geo}}\left(\pi-\tan ^{-1}(\sqrt{2}), \frac{\pi}{4},-\frac{2 \pi}{3}\right)
		$ & $ R_{\rm{non}}\left(\frac{3 \pi}{2}, 0, \frac{\pi}{2}, \frac{\pi}{2}\right)
		$\\
		\hline
		\hline
	\end{tabularx}}
	\label{ap:table}
\end{table*}

\bibliography{refs_GQG}

\end{document}